# SOHO CTOF Observations of Interstellar $He^+$ Pickup Ion Enhancements in Solar Wind Compression Regions

L. Saul[1], E. Möbius[1], Y. Litvinenko[1], P. Isenberg[1], H. Kucharek[1], M. Lee[1], H. Grünwaldt[2], F. Ipavich[3], B. Klecker[4], P. Bochsler[5]

*[1]University of New Hampshire, [2]Max Planck Institute for Aeronomy, [3]University of Maryland, [4]Max Planck Institute for Astrophysics, [5]University of Bern*

**Abstract.** We present a recent analysis with 1996 SOHO CELIAS CTOF data, which reveals correlations of $He^+$ pickup ion fluxes and spectra with the magnetic field strength and solar wind density. The motivation is to better understand the ubiquitous large variations in both pickup ion fluxes and their velocity distributions found in interstellar pickup ion datasets. We concentrate on time periods of that can be associated with compression regions in the solar wind. Along with enhancements of the overall pickup ion fluxes, adiabatic heating and acceleration of the pickup ions are also observed in these regions. Transport processes that lead to the observed compressions and related heating or acceleration are discussed. A shift in velocity space associated with traveling interplanetary compression regions is observed, and a simple model presented to explain this phenomenon based on the conserved magnetic adiabatic moment.

## INTRODUCTION

Although interstellar pickup ions (PUIs) stem from a supposedly stationary source of neutral gas, large variations of pickup He+ fluxes and their energy spectra in the solar wind have been observed on a variety of time scales. Some causes of these variations have been identified, including a depletion of pickup ions in the anti-sunward part of the distribution during time periods of radial interplanetary magnetic fields [3,7], variation of the ionization rate, and variation of the cutoff velocity of the He+ spectra due to the non-zero inflow speed of the neutral helium[1,8]. However, variations of the fluxes up to an order of magnitude, and substantial variations of the pickup ion spectra still remain unexplained. An understanding of these variations is necessary for using PUI measurements as probes of their source populations or for plasma transport parameters.

The CTOF instrument, in the CELIAS package on board SOHO, produced 150 days of data under relatively steady conditions in the upwind region of the interstellar flow. The relatively large geometric factor of the instrument, and it's location at the L1 point, make high time resolution observations of PUI spectra possible. This makes the dataset interesting for studying the short term variations in both total flux and velocity distribution.

The observations presented here were made during solar minimum, mostly in slow solar wind. The solar wind parameters used for this study are taken from the MTOF proton monitor on board SOHO. The magnetic field data are taken from WIND MFI, and convected back to the location of SOHO using the interplanetary magnetic field (IMF) orientation and measured solar wind speed. Our analysis of this dataset is in two parts. The first is a statistical analysis of the entire dataset, comparing PUI fluxes to solar wind parameters. The second analysis is of individual events in the dataset, for which enhancement of the PUIs in especially pronounced, either in the bulk or tail of the distribution. We look in detail at the evolution of the PUI distribution through these events.

## STATISTICAL ANALYSIS

By combining PUI events from time periods with specified solar wind conditions, and binning them in normalized velocity $V/V_{sw}$, spectra are formed that are representative of the chosen solar wind parameters. This method of statistical analysis can tell us not only whether the PUI flux and the solar wind parameters are correlated, but also where in the velocity spectrum the correlation is the strongest. We carried out this analysis with bulk solar wind parameters, including proton density, thermal velocity, and solar wind speed. We also carried out the analysis with IMF magnitude and direction.

The observed fluxes of the $He^+$ PUIs were found to be strongly correlated with the proton density (Fig. 1). The energy flux was always higher during periods of higher proton density for this analysis. A peak near the injection speed ($2V_{sw}$) is clearly visible in the spectra for $30cm^{-3} < n_p < 40cm^{-3}$. Also note that the peak energy of the ions is shifted toward the cutoff in higher density solar wind

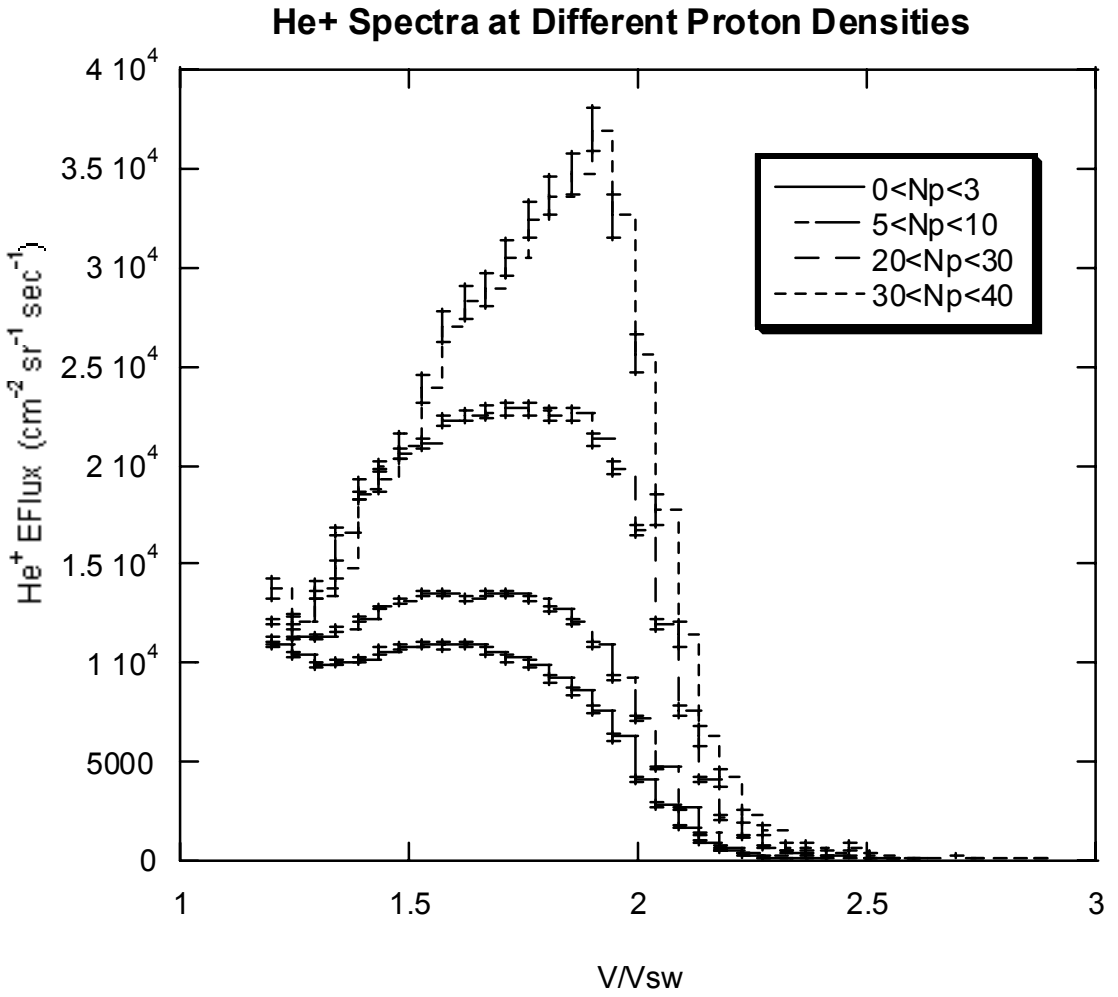

**FIGURE 1.** PUI differential energy flux spectra for four ranges of proton density in $cm^{-3}$. Error bars represent the statistical error.

Another strong correlation observed was with the IMF magnitude (Fig. 2). This complements previously observed correlations with the orientation of the IMF [3,7]. The PUI flux was found to not notably increase until |B| > 6nT, after which it increased monotonically with |B|. A peak near the injection speed was visible in the highest fields, similar to the peak in regions of highest proton density. We also performed the same analysis for restricted set of data based on IMF orientation. We found the correlation of PUI fluxes with |B| persisted for all orientations, becoming more pronounced as the field became more perpendicular to the solar wind velocity.

Because the PUI fluxes showed correlations with both IMF strength and field density, we then looked in more detail at times of solar wind compression, which have both enhanced IMF and proton density.

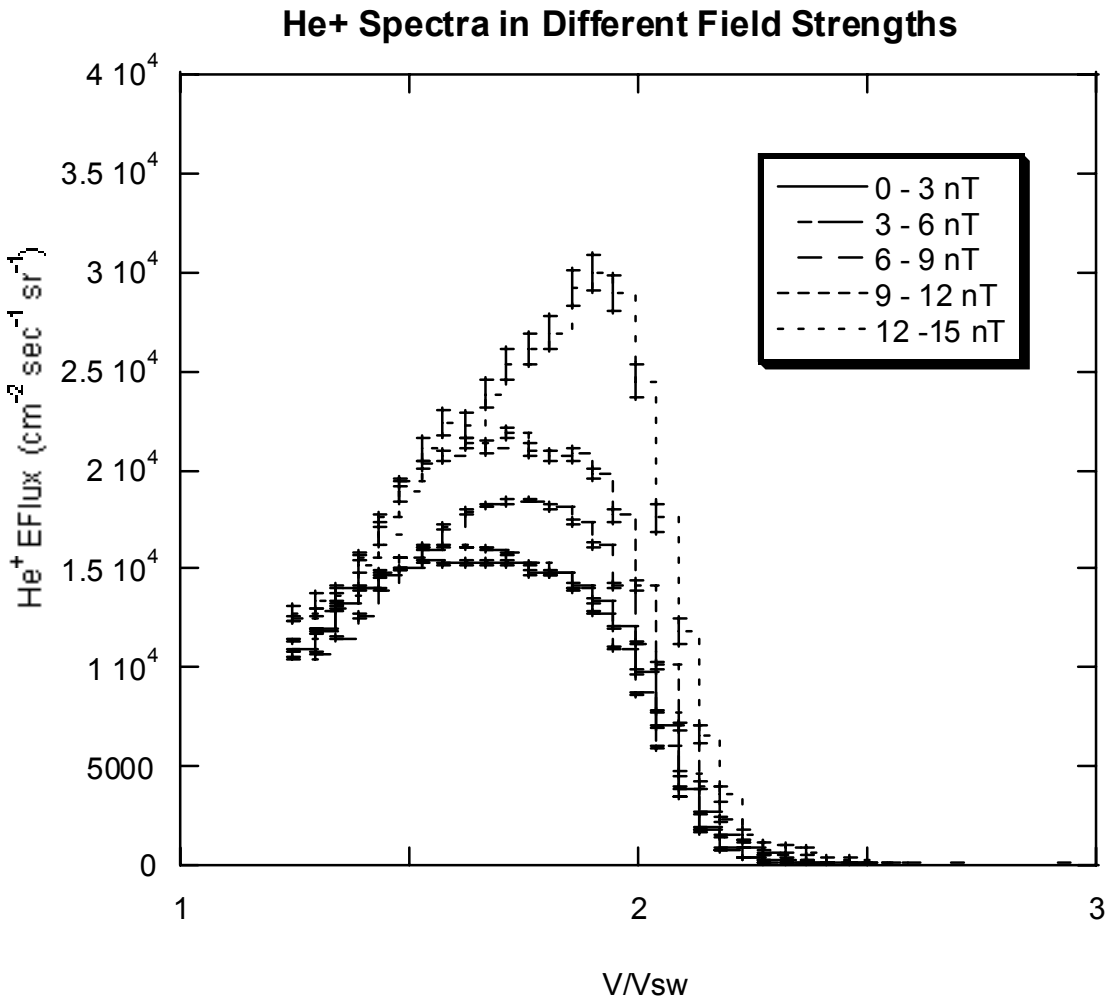

**FIGURE 2.** PUI differential energy flux spectra for five different ranges of field strength, integrated over the entire dataset.

## COMPRESSION REGIONS

SOHO saw several solar wind compression events during 1996. Other compression events in high Heliospheric latitudes have been shown to affect PUI fluxes [10]. Similar effects were observed here in the solar wind at 1AU in the ecliptic. Accelerated particles are observed in conjunction with the compression, especially downstream of the regions, here and elsewhere [9].

Some CIRs and transient stream interaction regions have also been observed over this time period with other instruments on SOHO [5]. We also observed the strong transient stream interaction of DOY 170 (Fig. 3), and a magnetic cloud CME around DOY 149. This cloud displayed compression of the solar wind and $He^+$ before and after its passage, with stronger compression trailing the region. The details of the PUI spectra varied in these compression regions, though the major features showed many similarities. The compression of the PUIs was strongest in the velocity range near the injection speed, analogous to the statistical correlations discussed above. Accelerated PUIs (last panel Fig. 3) were enhanced downstream of the regions. Some times with what looks like high PUI tail flux (e.g. high energy range near the magnetic discontinuity in Fig. 3) are actually due to a shift in velocity space of the distribution.

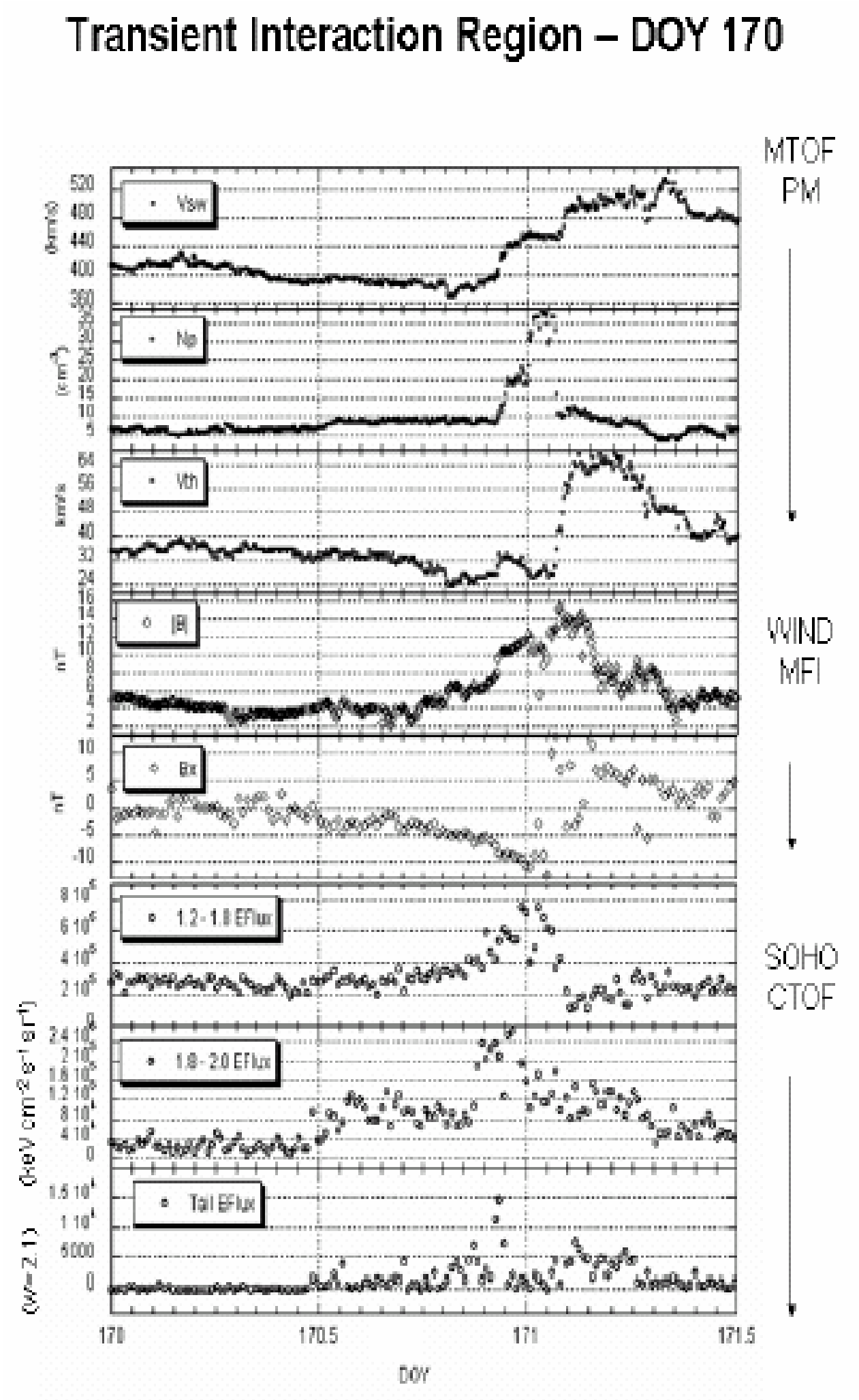

**FIGURE 3.** The stream interaction region above can be identified by the jumps in solar wind speed (1st panel), thermal speed (3rd panel), and the reversal of the radial component $B_x$ (5th panel). The PUIs are divided into 3 energy ranges in $w = V/V_{sw}$.

## Cutoff Shift at Magnetic Discontinuities

In some cases, the PUI spectra showed a shift in the cutoff velocity around the compression region (Figs. 3, 4). We hypothesized that this was due to conservation of $v_\perp^2/|B|$ as the distribution entered a compression region with enhanced IMF. Consistent with this explanation, the shift was only observed close to a jump in magnetic field magnitude. In addition, no shift was observed unless the discontinuity was traveling, i.e. fast wind pushing the discontinuity into slower wind.

This shift due to a conserved adiabatic moment will only affect perpendicular velocity. Therefore, the maximum PUI velocity shift is predicted for a ring distribution, for which the shift is: $v_2 = v_1\sqrt{B_2/B_1}$. To find the shift for an isotropic distribution we integrate over all pitch angles:

$$v_2 = v_1 \frac{\pi}{2}\int_0^{\pi}\sqrt{\frac{B_2}{B_1}\sin^2(\alpha)+\cos^2(\alpha)}\,\sin\alpha d\alpha\,,$$

which can be approximated as $v_2 \sim v_1\left(B_2/B_1\right)^{0.4}$. The initial velocity is taken (in the solar wind frame) to be the solar wind speed plus the radial component of the bulk neutral (Keplerian) velocity at the spacecraft's position in the Heliosphere [1,8].

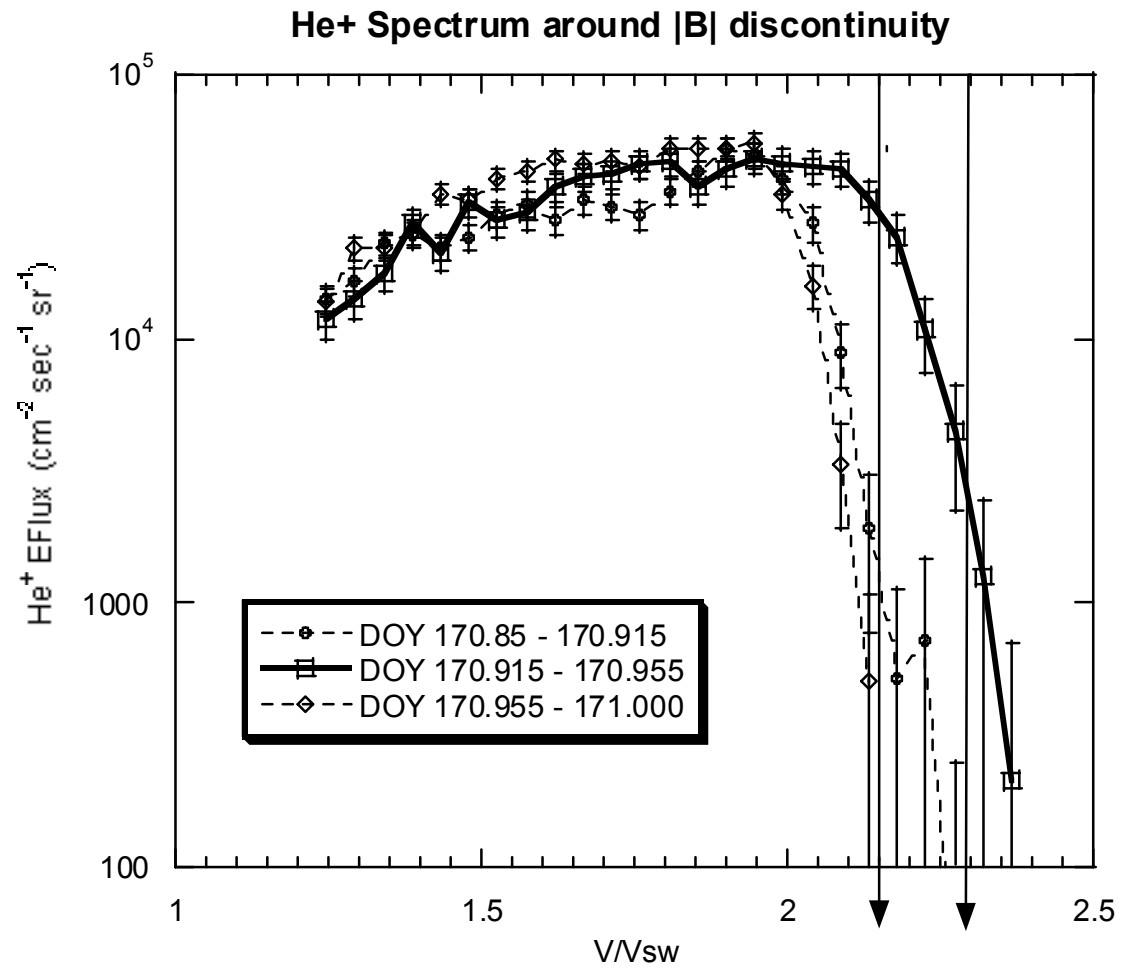

**FIGURE 4**. Logarithmic PUI differential energy flux spectra corresponding to three one hour time periods right at the magnetic discontinuity in Fig. 3.

The model described here for an isotropic distribution predicts the initial and compressed cutoff velocities shown in fig. 4 with downward arrows: $v_1$=2.15 and $v_2$=2.29 (for DOY 170, B2/B1 = 5/4, and assuming $V_{lism}$ = 27km/s).

To further test this interpretation the shift ratio $v_2/v_1$ was estimated using the inflection points of polynomial fits to the spectra (in linear representation). The results are shown for four compression events (fig. 5), including the one shown here at DOY 170, the trailing edge of the event at DOY 149, and two smaller compressions on DOY 99. The results are found to be consistent with the models. The exact type of distribution depends on the pitch angle scattering rate as the PUI distribution evolves. A real model of the spectra (rather than a polynomial fit) would be needed to differentiate between the different types of distributions.

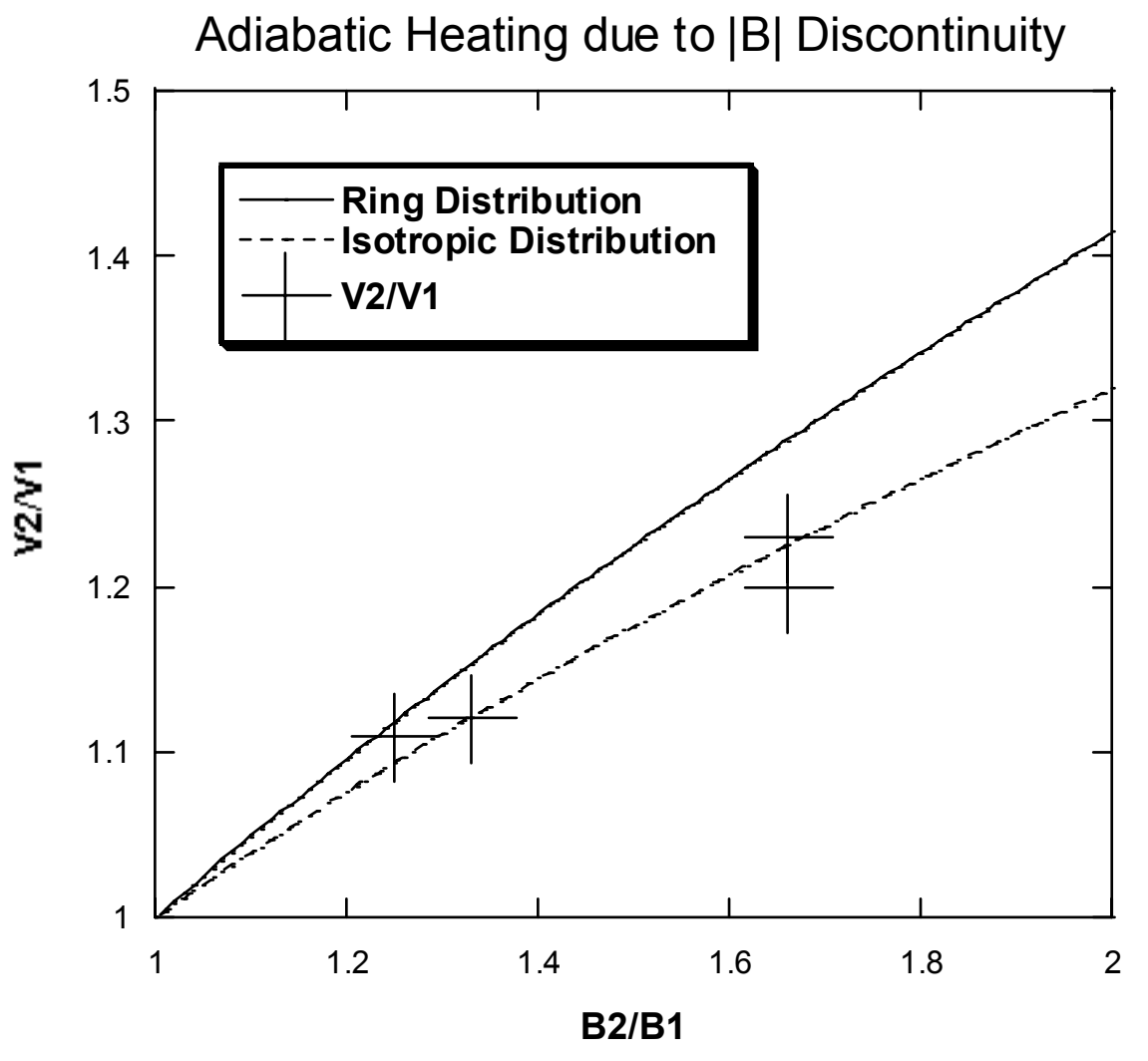

**FIGURE 5**. Predicted velocity shift ratios (lines) for isotropic and ring distributions are shown vs. the corresponding magnetic field compression ratio. The estimated shifts in 4 observed distributions are also shown.

## DISCUSSION

The CTOF measurements show strong correlations with solar wind parameters, especially the proton density. This is perhaps not a surprise; one would expect conditions that compress the bulk solar wind to compress the PUIs contained in the wind. The observed correlation with the IMF magnitude was similar in character to that of the proton density, pointing to a similar cause for these two correlations. The observations are consistent with the hypothesis that compression regions are responsible for the many of the PUI flux variations.

The compressions not only enhance the overall flux, but change the shape of the distribution, most notably in an enhancement near the injection velocity in compressed regions. The observations are consistent with the compression compensating for some of the adiabatic cooling of the PUIs, thus keeping their speeds in the area around 2Vsw that would have cooled in the expanding wind. However, strong shocks are known to affect PUI distributions in other ways, including heated electrons increasing the ionization rate [6]. It could be that something similar takes place in compression regions, contributing to the enhancement near the cutoff.

The observed velocity shift due to magnetic discontinuities can be explained with adiabatic effects. Further work is needed to model it more fully and interpret its influence on different distributions or in different solar wind conditions.

## ACKNOWLEDGMENTS

This work is partially supported by NASA grant NGT5-50381, NAG5-10890, and NST ATM-9800781. Thanks to the WIND/MFI team for the IMF data.